\documentclass{article}

\usepackage[preprint]{neurips_2021}

\usepackage[utf8]{inputenc} 
\usepackage[T1]{fontenc}    
\usepackage{hyperref}       
\usepackage{url}            
\usepackage{booktabs}       
\usepackage{amsfonts}       
\usepackage{nicefrac}       
\usepackage{microtype}      
\usepackage{xcolor}         

\usepackage{caption}
\usepackage{subcaption}
\usepackage{graphicx}
\usepackage[title]{appendix}

\title{Towards better data discovery and collection with flow-based programming}

\author{%
	Andrei Paleyes \\
	Department of Computer Science\\
	University of Cambridge\\
	\texttt{ap2169@cam.ac.uk} \\
	\And
	Christian Cabrera \\
	Department of Computer Science\\
	University of Cambridge\\
	\texttt{chc79@cam.ac.uk} \\
	\And
	Neil D. Lawrence \\
	Department of Computer Science\\
	University of Cambridge\\
	\texttt{ndl21@cam.ac.uk} \\
}

\begin{document}

\maketitle

\begin{abstract}
Despite huge successes reported by the field of machine learning, such as voice assistants or self-driving cars, businesses still observe very high failure rate when it comes to deployment of ML in production. We argue that part of the reason is infrastructure that was not designed for data-oriented activities. This paper explores the potential of flow-based programming (FBP) for simplifying data discovery and collection in software systems. We compare FBP with the currently prevalent service-oriented paradigm to assess characteristics of each paradigm in the context of ML deployment. We develop a data processing application, formulate a subsequent ML deployment task, and measure the impact of the task implementation within both programming paradigms. Our main conclusion is that FBP shows great potential for providing  data-centric infrastructural benefits for deployment of ML. Additionally, we provide an insight into the current trend that prioritizes model development over data quality management.
\end{abstract}

\section{Introduction}\label{section:intro}
After achieving considerable success as an academic discipline, machine learning (ML) components are now increasingly deployed in production systems. Government agencies, private companies and individuals all apply ML to solve practical tasks. McKinsey has reported a 25\% year to year growth of ML adoption by businesses, with nearly half of respondents reporting revenue increase \citep{cam2019global}.

Deployment of machine learning typically happens on top of the existing data processing infrastructure. Companies aim to speed up their processing workflows, gain additional insights to aid their decision making, improve detection of anomalous behavior, or provide customers with new functionality based on historical data. Machine learning models are often a centerpiece of such projects. Unfortunately, ML deployment projects face difficult challenges, with companies reporting up to 50\% failure rate \citep{wiggers2019idc}. We believe that one of the reasons for these failures lies in the fact that the majority of ML projects are being deployed on top of existing software solutions which were built to fulfill goals that are important but unrelated to ML, such as high availability, robustness, low latency. However, machine learning poses a new set of challenges that the majority of existing software architectures are not designed for \citep{paleyes2020challenges}. Data processing is mentioned as one of the areas that cause most concern \citep{polyzotis2018data}, especially in high-scale service-oriented software environments, such as Twitter \citep{lin2013scaling} or Confluent \citep{stopford2016data}.

As more businesses seek to convert the data they manage into value, it seems reasonable to explore software architectures that could better fit that purpose. In this work we consider the potential of flow-based programming (FBP, \cite{morrison1994flow}) as a paradigm for building business applications with ML, and compare it with currently prevalent control-flow paradigms, namely service-oriented architecture (SOA, \cite{perrey2003service, oreilly2020microservices}). 

There have already been some attempts to enhance SOA with better data handling capabilities \citep{gotz2018challenges, gluck2020doma, safina2016data, dehghani2019move}. In our work we wanted to explore radically different approach towards better data management in software systems. So, rather than incrementally improving SOA, we consider FBP due to a range of useful properties that are particular to the paradigm, e.g. access to a dataflow graph and data coupling between the components. We anticipate that they make data-related tasks, such as data discovery and collection, simpler to perform. We illustrate this idea with a simple experiment. We develop an example application separately with each paradigm. We then carry out an ML deployment procedure within both implementations, and analyze how each deployment stage affects the complexity of the codebase. Our conclusions show that while there are a number of trade-offs to consider, FBP has potential to simplify deployment of ML in data-driven applications.

Data flow paradigms are not new in software engineering, a duality of control flow and data flow for building software has long been explored by the computer science community \citep{treleaven1982towards, lauer1979duality, hasselbring2021control}. FBP's potential to improve software quality and maintenance has been shown in comparison with other paradigms and design principles, such as OOP \citep{morrison2010flow}, functional programming \cite{roosta2000data}, and SOA in IoT context \citep{lobunets2014applying}. But to our knowledge FBP has never been compared to SOA in the context of ML deployment before. Some of the high level ideas that motivated this paper were first introduced by \cite{diethe2019continual} and further developed by \cite{lawrence2019doa} and \cite{borchert2020milan} under the name of Data Oriented Architectures (DOA). Our work can be seen as the first step towards applying DOA to practical tasks.

\section{Experiment setup}

To explore the different software paradigms we developed an example application to study properties of FBP and compare it against the more widespread SOA approach. Concretely, we implemented a prototype of a taxi ride allocation system described by \cite{lawrence2019doa}. The application receives data about currently available drivers and incoming ride requests, and outputs the allocated rides. The application also processes updates of each allocated ride, and keeps track of factual passenger wait times. We formulated a business problem that can be solved with ML: provide user with an estimated wait time, in addition to the allocated driver. Training data for the ML model can be collected based on historical wait times. This type of additional functionality has been shown to be among the major contributors to project's technical debt \citep{molnar2020long}. As a result we focus our evaluation on changes in code quality.

Two separate implementations of the described application were created: one with FBP using flowpipe\footnote{flowpipe is available at https://github.com/PaulSchweizer/flowpipe. It is considered to be an FBP-inspired framework, but provides all FBP features critical for our work and is easy to read and understand.} and one with SOA using Flask\footnote{https://flask.palletsprojects.com/}. Detailed description of the application and implementation details specific to each paradigm can be found in Appendix~\ref{appendix:application}. Full source code of the application can be found at \url{https://github.com/mlatcl/fbp-vs-soa/tree/ride-allocation}.

In order to enable structured approach towards evaluation of codebase changes over the course of ML deployment, we defined three stages of the implementation:
\begin{itemize}
	\item \textbf{Stage 1}: minimal code to provide basic functionality. The stage is denoted in the code and this paper by suffix \textit{min}.
	\item \textbf{Stage 2}: same as Stage 1 plus dataset collection. A complete dataset required collecting data from two locations within the application. Inputs, which we considered to be ride requests and driver locations, are available at the time ride allocation is done. Output, which is the actual waiting time, becomes available later in different part of the app, when passenger pickup happens. Denoted by suffix \textit{data}.
	\item \textbf{Stage 3}: same as Stage 2 plus the new output of estimated wait time produced via a deployed ML model. The ML model is trained on the dataset collected at the previous stage. The application has to load an already trained offline and serialized ML model, perform the prediction at the time ride allocation is done, and output estimated wait time in addition to the allocation information. Denoted by suffix \textit{ml}.
\end{itemize}

Overall we ended up with 6 versions of the Ride Allocation system, which are listed in Table~\ref{table:app_list}.

\begin{table*}[t]

	\caption{List of all created versions of the Ride Allocation app. First column gives the key by which a particular version is referred to in the codebase. Explanation of each metric used can be found in Appendix~\ref{appendix:evaluation}.}
	\label{table:app_list}
	\begin{center}
		\begin{tabular}{ |l|l|l|l| }
			\hline
			\textbf{Key} & \textbf{Paradigm} & \textbf{Stage} & \textbf{Description} \\
			\hline
			$fb\_app\_min$ & FBP & 1 & Basic functionality \\
			\hline
			$fb\_app\_data$ & FBP & 2 & Same as $fb\_app\_min$ plus dataset collection \\
			\hline
			$fb\_app\_ml$ & FBP & 3 & Same as $fb\_app\_data$ plus estimated wait time output \\
			\hline
			$soa\_app\_min$ & SOA & 1 & Basic functionality \\
			\hline
			$soa\_app\_data$ & SOA & 2 & Same as $soa\_app\_min$ plus dataset collection \\
			\hline
			$soa\_app\_ml$ & SOA & 3 & Same as $soa\_app\_data$ plus estimated wait time output \\
			\hline
		\end{tabular}
	\end{center}

\end{table*}

We use a number of software metrics to assess the impact of each subsequent stage on the overall quality of the codebase. These metrics measure size, complexity and maintainability of the code. This metric-based evaluation approach was chosen to enable objective evaluation of codebase quality and how the ML deployment process affected it at each stage. 

\section{Experiment Results}

In this section we discuss our observations from the experiment, summing up our observations from each stage. Detailed discussion of the evaluation, as well as our subjective impressions of those development aspects that were harder to measure, can be found in Appendix~\ref{appendix:evaluation}.

Size and complexity metrics and our own observations suggest that initial cost of developing the FBP solutions is higher. That is likely the consequence of the fact that SOA is a highly evolved and widely deployed programming paradigm. Therefore the majority of people with experience of modern industrial software development, which authors of this work consider themselves to be, can iterate and make progress within this paradigm at a quicker pace. In contrast, FBP is not nearly as widespread. This paradigm requires a conceptual shift in the way a developer thinks about the application, because instead of customary control-flow one needs to adopt data-flow mindset. However, cognitive complexity metric suggest that FBP programs are easier to read and comprehend once they are written (see Figure~\ref{figure:cognitive_metrics} in the Appendix~\ref{appendix:evaluation}).

Dataset collection stage turned out to be the most critical for surfacing differences between the paradigms in the ML deployment context. FBP programs allow programmatic access to the whole dataflow graph, with nodes representing business logic of the application, and edges representing flows of data, meaning that it is possible to programmatically access data flowing to or from any node. Consequently, changes were made in single location of the codebase, even though we needed to collect data from multiple internal sources. In contrast, changes to the SOA application had to be introduced in multiple places, which poorly affects code intepretability and introduces challenges for long term support.

Unlike the previous two stages, the model hosting stage yielded no additional insight into difference between the paradigms considered. Nevertheless it is important to see the confirmation of the fact that both paradigms can support hosting ML model for predictions without significant impact on the rest of the system.

\begin{figure*}[ht]
	\captionsetup[subfigure]{width=0.8\linewidth,justification=raggedright}
	\begin{subfigure}{.45\textwidth}
		\centering
		\includegraphics[width=\linewidth]{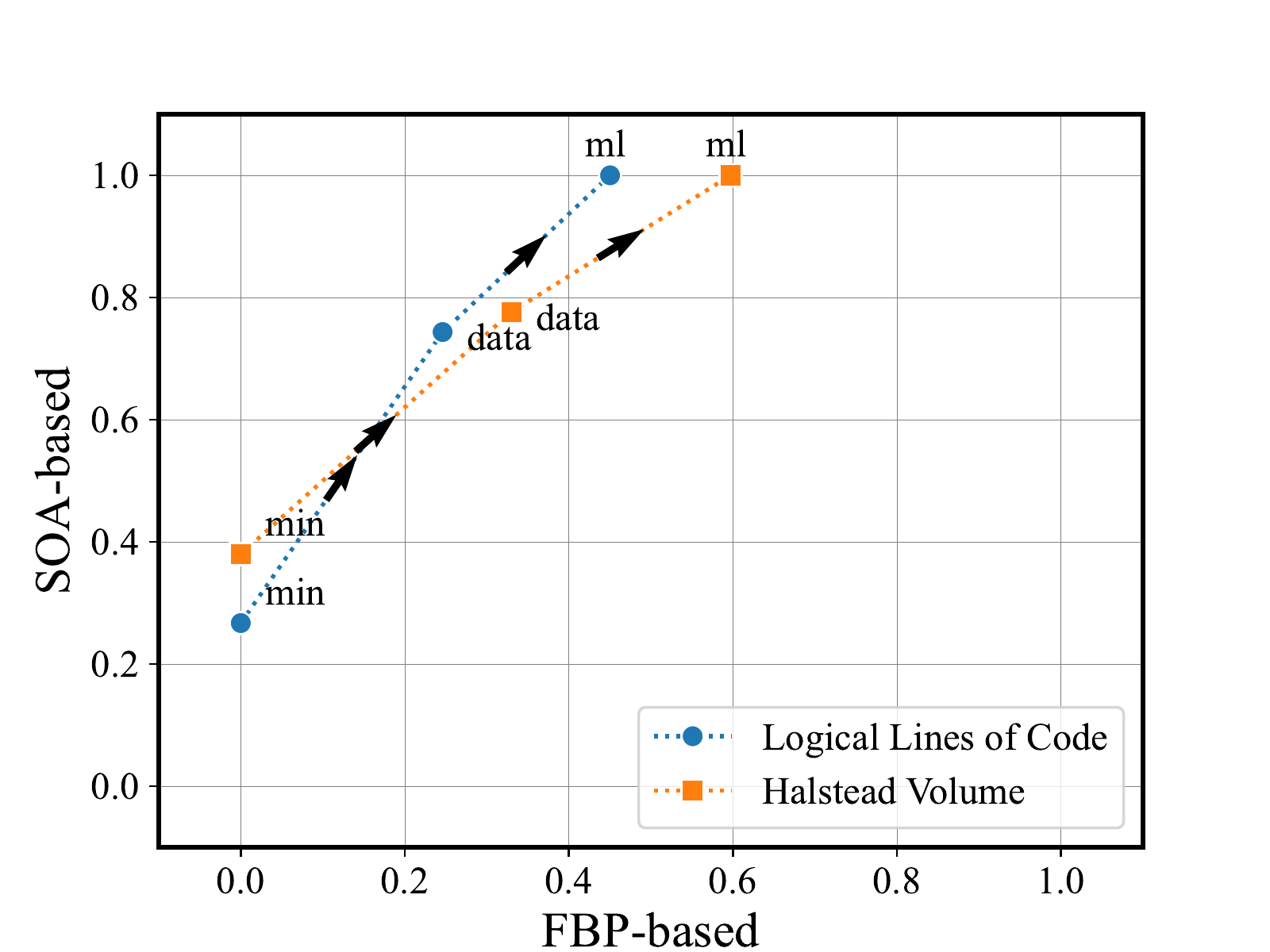}
		\caption{Code size metrics combined. The trends intersect between stages \textit{min} and \textit{data}, while being nearly parallel between stages \textit{data} and \textit{ml}. This suggests that data collection on \textit{data} stage had different scale of impact on metrics for FBP and SOA implementations.}
	\end{subfigure}
	\begin{subfigure}{.45\textwidth}
		\centering
		\includegraphics[width=\linewidth]{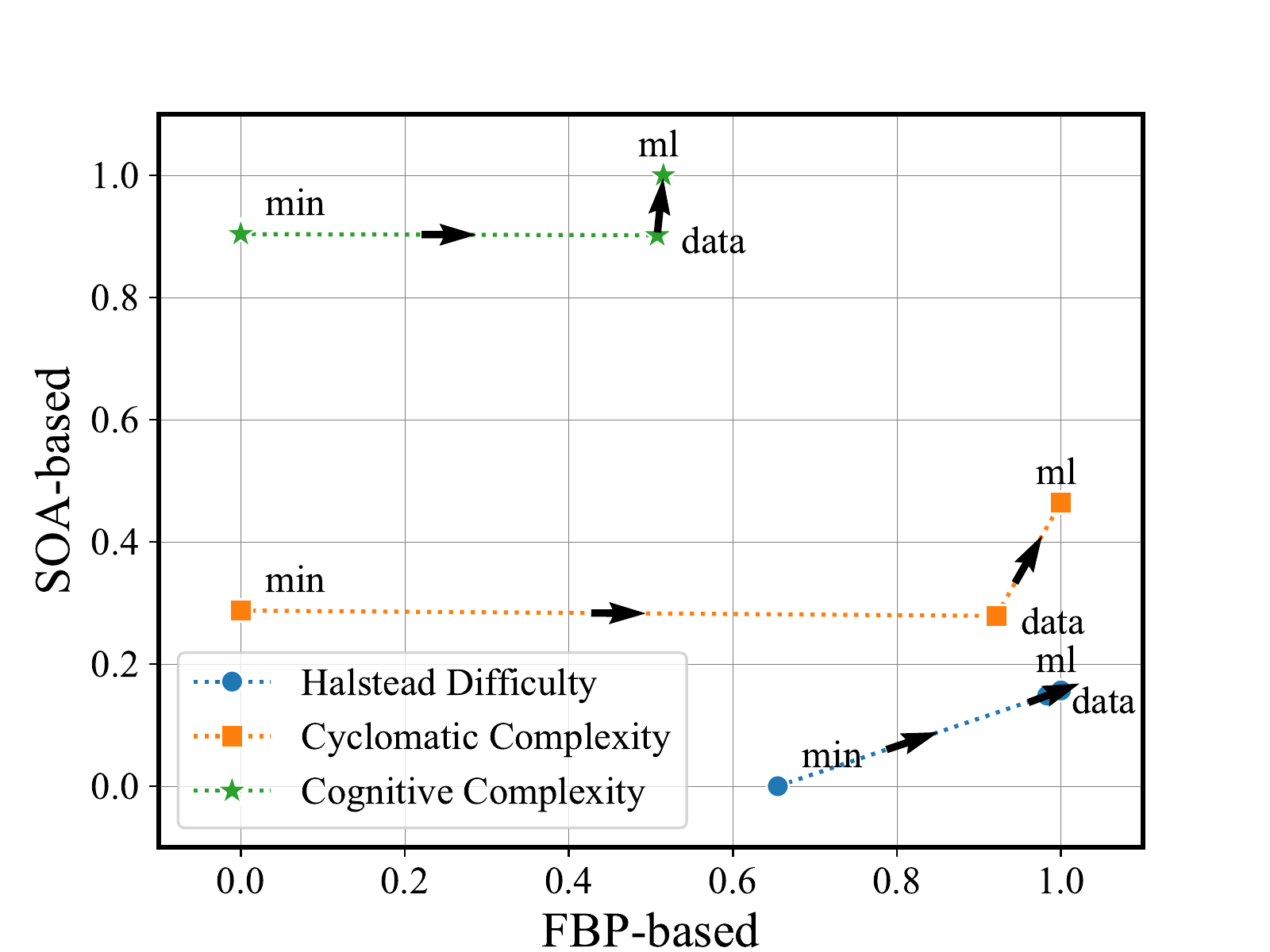}
		\caption{Code complexity metrics combined. The distance between points that represent stages \textit{min} and \textit{data} is significantly bigger than between points for stages \textit{data} and \textit{ml}. This shows how big was an impact on code complexity by the implementation of data collection.}
	\end{subfigure}
	\caption{Combinations of multiple code metrics. All values are normalized to fall within $[0, 1]$ range. Text next to a point is the app version key this point corresponds to.}
	\label{figure:combined_metrics}
\end{figure*}

Comparing behavior of multiple code complexity metrics we have realized that the data collection stage was far more impactful change than the model deployment  (Figure~\ref{figure:combined_metrics}). This might uncover additional reason for the trend in modern ML community to focus on model research rather then data research \citep{lawrence2017data}. If making changes to deployed model is easier and less error-prone than making changes to data engineering pipeline, it is easy to understand why developers and researches are motivated to seek improvements in model iterations rather than over data quality. Nevertheless we believe data management is equally important part of machine learning process, especially since data scientists spend most of their time working with data \citep{nazabal2020data}. With this work we aim to make a step towards simplifying data-oriented tasks in software systems.

\section{Conclusions and Future Work}

In this paper we illustrated the potential of using FBP to ease the pain of deploying ML and improve data management. In a software system designed according to FBP principles the tasks of data discovery and collection become more straightforward, thus simplifying consequent deployment of ML. We believe better tooling that allows developers to define dataflow graphs at a higher level of abstraction would help fill some of the current gaps and leverage that potential.

Additionally, we showed that data collection code caused much more significant impact to metrics of both codebases, compared to model deployment. This could be an explanation of modern trends of seeking performance improvements through models rather than through data.

We observed that when developing an application with FBP paradigm, a lot of effort is spent in defining and manipulating the dataflow graph. On the other hand, once such a graph is defined, all data flows in the system become explicit, thus making data discovery task simpler. Any framework that allows software developers to abstract away from the boilerplate code and focus on actual application domain, business logic and entities, would streamline the development process and reduce the complexity of the codebase. There are tools that might serve this purpose, such as Google Dataflow\cite{krishnan2015google}, Kubeflow\cite{bisong2019kubeflow} and Apache NiFi\footnote{https://nifi.apache.org/}, although they are usually seen as very specific to particular applications. Understanding commonalities of these frameworks is a promising starting point for building general purpose development tools.

In the future we would like to further scale the experiment described in this paper. For instance, the same ML deployment perspective can be considered in the distributed context, where data streaming platforms such as Apache Kafka would have to be used. Long-term experiments can also be informative to observe the code evolution over a longer period, e.g. a year. Other paradigms, such as the Actor model \citep{hewitt2010actor}, might be considered for comparison.

\begin{ack}
We would like to thank our colleagues Pierre Thodoroff, Markus Kaiser, Eric Meissner, Jessica Montgomery, Diana Robinson for many insightful discussions.
\end{ack}

\bibliographystyle{unsrtnat}
\bibliography{references}


\begin{appendices}

\section{Background on relevant paradigms}\label{appendix:background}

In this section we briefly introduce key concepts and paradigms used throughout the paper.

\subsection{Service-oriented architecture}
Service-oriented architecture (SOA) is a paradigm for development of software applications \cite{perrey2003service}. Under this paradigm the application is broken down into several components called services, which interact between each other via a pre-defined protocol to provide application's users with necessary functionality. Services only interact on API level, thus hiding details of each service's implementation, which results in a set of loosely coupled components. In recent years SOA and its derivatives, such as microservices, gained substantial popularity and can be considered de-facto standard distributed software design approach \cite{oreilly2020microservices}. Service orientation gives developers a range of important benefits: incapsulation, loose coupling, modularity, scalability and ease of integration with other systems, so it is a reasonable choice for those who need to build scalable software~\cite{papazoglou2003service,cabrera2017implementing}.

\subsection{Flow-based programming}
Flow-based programming was created by \cite{morrison1994flow}, and can be considered a special case of the more general dataflow programming paradigm. It defines software applications as a set of processes which exchange data via connections that are external to those processes. FBP exhibits ``data coupling'', which is considered in computing to be the loosest form of coupling between software components, thus promoting a flexible component-oriented software architecture. FBP has a reputation as a paradigm that optimizes speed, bandwidth and processing power in multi-tasking, parallel processing applications. For example Szydlo et. al. consider FBP's application to IoT \cite{szydlo2017flow}, Lampa et.al. explored FBP's potential in the context of drug discovery \cite{lampa2016towards}, Zaman et. al. presented an FBP-based framework for mobile development \cite{zaman2015flow}. Recent years saw birth of several general-purpose projects built on flow-based programming principles, such as NoFlo \cite{bergius2015noflo} and Node-RED\footnote{\url{https://nodered.org/}}. Node-RED in particular became popular in the IoT community \cite{clerissi2018towards, chaczko2017learning}, as the FBP model was found to be a good fit for building data processing pipelines in IoT. Developing this idea further, in this paper we argue that there is potential in a wider use of FBP.

A notable feature of FBP is the ability to present the whole program visually as a graph of data flow. This feature has two important implications. First, the graph-like structure allows to reason about the complete programs in a unique way that is often impossible in case of object oriented programming (OOP) or SOA. We leverage this aspect in this work. Second, it allows for visual no-code programming which in some cases may aid adoption of FBP. In particular it is believed to be useful for beginners who have little or no prior coding experience \citep{mason2017block}.

\subsection{Data streams}
In this work we make use of data streams as connectors in flow-based programs. A data stream is a sequence of data records that are made available over time. Machine learning on data streams is not a new concept. Data processing platforms such as Apache Spark \citep{meng2016mllib}, Apache Flink \citep{carbone2015apache} or Google Cloud Dataflow \citep{krishnan2015google} are widely used for manipulating large datasets and executing machine learning tasks. AWS Kinesis\footnote{\url{https://aws.amazon.com/kinesis/}} and Apache Kafka \citep{kreps2011kafka} are some of the most commonly used data streaming services.

\section{Ride Allocation application}\label{appendix:application}

In this section we discuss the gory details of implementing Ride Allocation application in SOA and FBP paradigms, starting with the general description of the application itself.

\subsection{Ride Allocation application}\label{subsection:ride_allocation_system}
We have chosen Ride Allocation application as described by \cite{lawrence2019doa} for our case study. On one hand, the idea of assigning taxi drivers to incoming ride requests is simple enough to describe and it does not require much background knowledge. On the other hand, it provides enough non-trivial functionality and data processing to make it worth studying in the context of ML deployment.

\begin{figure}[ht]
	\centering
	\includegraphics[width=0.7\columnwidth]{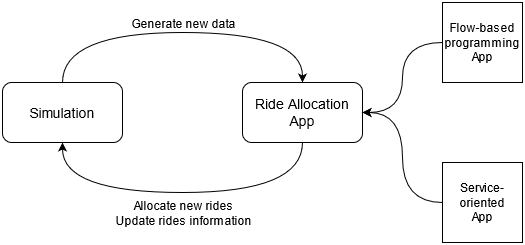}
	\caption{Overview of the Ride Allocation system high level mechanics. Simulation part generates new data, which is handled by Ride Allocation app to allocate new rides or update information about existing ones. Output of the app is sent back to the simulation. All FBP and SOA implementations of the Ride Allocation app implement same interface to ensure fair comparison.}
	\label{figure:app_diagram}
\end{figure}

At the highest level our implementation consists of two parts: the allocation system itself and the code that simulates events happening in the outside world. This overview can be seen in Figure~\ref{figure:app_diagram}.

The allocation system provides several functions. First, it assigns taxi drivers to incoming ride requests. Second, it keeps track of all allocated rides and updates them according to the incoming events. We recognize several types of events: ride starts, ride finishes, location updates, cancellations. Third, it calculates factual wait times for passengers, defined as a difference between the moment a ride was allocated and the passenger's pickup.

The simulation part is implemented as a discrete-event simulation and is responsible for generating events that would be happening if our system was deployed as a part of a real life taxi application. In addition to creating initial data like a list of available drivers, it generates new ride requests and events for rides that were previously allocated.

\begin{figure}[ht]
	\centering
	\includegraphics[width=\columnwidth]{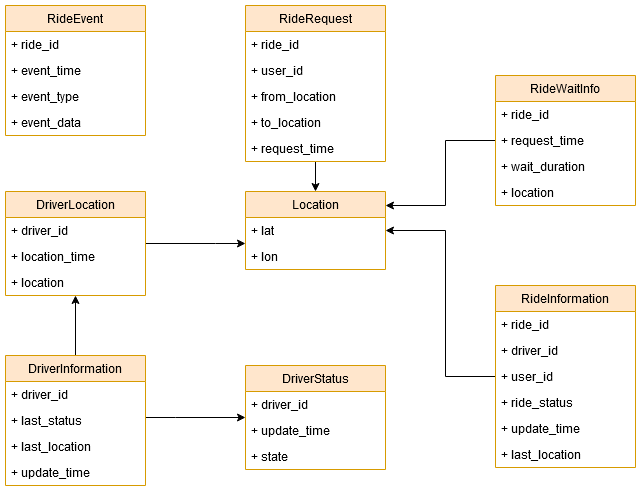}
	\caption{All data record types defined in the Ride Allocation system. Each table shows the name of the class and a list of its fields. Arrows signify "has a" relationship between entities. For simplicity separate types for driver and user were not defined, and instead just IDs were used to distinguish these entities. Although such types would be necessary in real software application, in our example they are not needed for any function.}
	\label{figure:record_types_diagram}
\end{figure}

To ensure fair comparison we have created functionally equivalent systems using FBP and SOA development paradigms. To further ensure equivalency we have maximized code reuse. In particular, both implementations share the same data types as well as the simulation code. All data types used in the system are shown in Figure~\ref{figure:record_types_diagram}.

\subsection{Implementation notes: SOA paradigm}

\begin{figure}[ht]
	\centering
	\includegraphics[width=0.7\columnwidth]{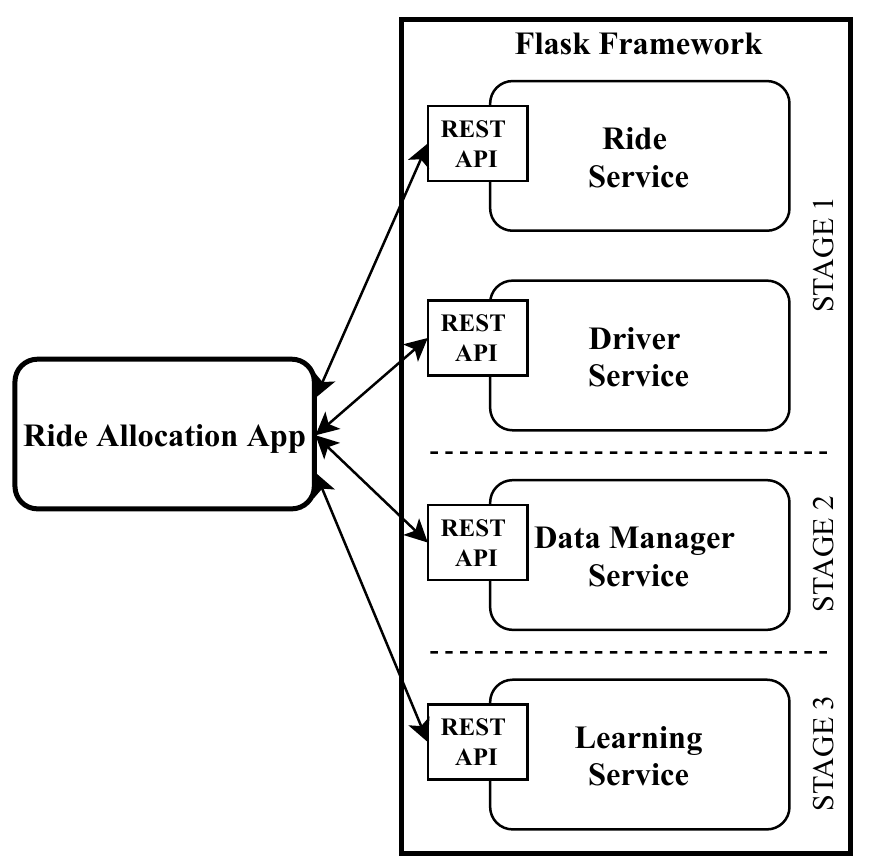}
	\caption{SOA implementation of the Ride Allocation app. New services defined within each stage are shown. App code interacts with each service via HTTP requests.}
	\label{figure:soa_diagram}
\end{figure}

Figure~\ref{figure:soa_diagram} shows the software implementation of the allocation app based on the SOA paradigm. We use the lightweight Flask\footnote{Flask Framework - \url{https://palletsprojects.com/p/flask/}} framework to develop the SOA based applications because of its flexibility and popularity among the micro-services community. We implement a set of services that offer the required capabilities for each implementation stage. The Ride Allocation App communicates with the services using REST APIs to save, retrieve and process rides, drivers, and waiting times data. All services are hosted locally, and the communication is happening via HTTP requests.

Two services were implemented for the Stage 1 as they provide the basic functionalities of the Ride Allocation application.

\textbf{Ride Service} implements the capabilities to save, retrieve, and process rides information according to the simulation events. Three APIs are implemented to store the rides information (i.e., \textit{add\_all\_rides}, \textit{add\_ride\_events}, and \textit{add\_ride\_infos}). Four APIs are implemented to retrieve rides information (i.e., \textit{get\_ride\_infos}, \textit{get\_ride\_wait\_infos}, \textit{get\_all\_rides}, and \textit{get\_rides\_by\_state}). And the \textit{update\_ride\_allocation} API is implemented to update rides' information once they are allocated.

\textbf{Driver Service} implements the capabilities to save, retrieve and process drivers information. We implement the \\ \textit{add\_all\_drivers\_statuses} and the \textit{add\_ride\_drivers\_locations} APIs to store drivers information. Three APIs are implemented to retrieve drivers data (i.e., \textit{get\_all\_drivers}, \textit{get\_driver\_data\_to\_save}, and \textit{get\_driver\_by\_id}). The \textit{allocate\_driver} and the \textit{release\_driver} APIs are implemented to update drivers availability once a ride is assigned or finished.

The services above fulfill all functionality required by Stage 1 of the experiment. For Stage 2 we implemented an additional service responsible for collecting the dataset. Specifically, \textbf{Data Manager Service} implements the functionality of collecting the data that the learning model uses in the next stage. These functions include data formatting (i.e., \textit{get\_ride\_data\_to\_save}, and \textit{get\_wait\_times\_to\_save}) and data storing (i.e., \textit{save\_data\_to\_file}). In addition, changes were made to \textbf{Ride} and \textbf{Driver} services to retrieve necessary data.

One more service was implemented in order to host the trained ML model and do predictions with it, as required by Stage 3 of the experiment. \textbf{Learning Service} provides the API \textit{get\_estimated\_times}, which predicts user's waiting times based on the driver's and user's locations using the trained model.

Furthermore, we have also implemented data access layer (DAL) that stored and retrieved data as necessary from the local SQLite database. For the purposes of fair comparison we have excluded DAL code from further analysis presented in Section~\ref{appendix:evaluation}, since FBP implementation described in the next section does not use a database.

\subsection{Implementation notes: FBP paradigm}
The FBP implementation of the Ride Allocation system is built with two major building blocks: data streams and stateless processing nodes. It follows an example implemented in the Milan package\footnote{\url{https://github.com/amzn/milan/tree/master/milan/milan-samples/milan-sample-gen/src/main/scala/com/amazon/milan/samples/bodaboda}}. For the purposes of this work we use the lightweight Python FBP framework flowpipe\footnote{\url{https://github.com/PaulSchweizer/flowpipe}}.

A data stream is a collection of data records of the same type that is updated whenever a new record arrives. Each data stream within the application belongs to one of three categories:
\begin{itemize}
	\item Input streams, that receive data from the outside world;
	\item Output streams, that hold data produced by the application;
	\item Internal streams, that hold intermediate data within the application. These streams are necessary because processing nodes by definition are not allowed to have state.
\end{itemize}

A processing node takes one or more streams as an input, performs some operations on them, and then puts the result into one or more output or internal data streams. Within the design approach we are considering such nodes do not carry internal state, and do not make external calls to outside services or databases. All data influencing a processing node should be registered in the system, so if such additional input is necessary, it should be represented as a data stream.

\begin{figure*}[ht]
	\centering
	\includegraphics[width=\textwidth]{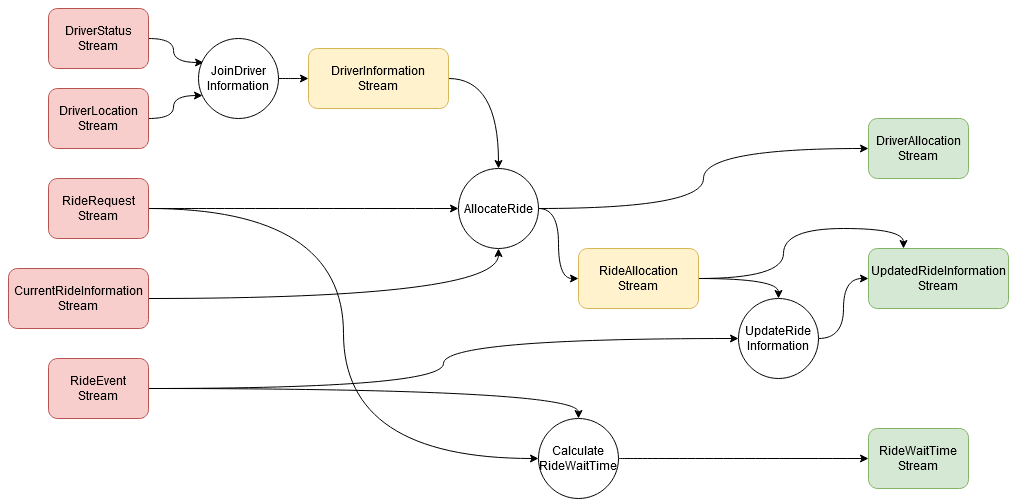}
	\caption{Flow diagram of the Ride Allocation system implemented via flow-based programming paradigm. Here rectangles show data streams, circles show processing nodes, arrows show flow of data records. Left-most streams (highlighted red) represent input data, right-most streams (green) represent system outputs, and other streams (yellow) are internal to the system.}
	\label{figure:fbp_flow_diagram}
\end{figure*}

As can be seen in Figure~\ref{figure:fbp_flow_diagram}, our system defines four granular operations, some of which produce data that is only observed within the application itself, such as joined information about a driver. This data is put on a separate stream before being consumed by the \textit{AllocateRide} operation.

Implementation of the dataset collection task in Stage 2 exploits the graph-like nature of FBP programs. In order to gain access to necessary data records, we traversed the dataflow graph and intercept data in corresponding nodes.

New output required by Stage 3 was implemented as one more output stream preceded by a new processing node. The processing node \textit{EstimateRideWaitTime} consumed \textit{RideAllocationStream} as well as \textit{DriverInformationStream}, loaded trained ML model to do necessary inference, and produced \textit{EstimatedRideWaitTime} output stream.

\section{Evaluation}\label{appendix:evaluation}
We employed two approaches towards evaluating pros and cons of each of the programming paradigms used in our experiment: metric-based and empirical comparison.

There are many possible aspects to consider when comparing software engineering approaches. Even though it is possible to draw some general conclusions about each paradigm from the comparisons outlined below, within the scope of this paper we only limited ourselves to the task of comparing FBP and SOA in the context of data management and ML. Therefore in this section we mostly consider visible impact on the overall codebase caused by implementation of tasks required by Stages 2 and 3 of the project.

\subsection{Metric-based comparison}
We use a number of software evaluation metrics in order to measure the impact of each subsequent stage on the overall quality of the codebase. Here is the complete list of metrics we used, with brief discussion of why we felt it is relevant to our experiment. We used the Python package Radon\footnote{\url{https://github.com/rubik/radon}} and the Flake8 plugin\footnote{\url{https://github.com/Melevir/cognitive_complexity}} to carry out measurements.

\begin{itemize}
	\item \textbf{Logical Lines of Code}, which only counts executable statements and ignores comments and blank lines, as a simple first measure of the codebase size.
	\item \textbf{Halstead Volume} \cite{halstead1977elements}, which defines code volume via the number of operations and operands. This is used as a more advanced measure of the codebase size.
	\item \textbf{Cyclomatic Complexity} \cite{mccabe1976complexity}, which measures the number of independent paths through the code. We use it to assess how the complexity of the codebase grows with added functionality. Since cyclomatic complexity is measured separately for each code block, such as a function, we consider average cyclomatic complexity across the application.
	\item \textbf{Cognitive complexity} \cite{campbell2018cognitive}, which measures the complexity of the control flow of code. We use it to assess how easy it is to understand the code. As with cyclomatic complexity, we consider average cognitive complexity across the application.
	\item \textbf{Halstead Difficulty} \cite{halstead1977elements}, which measures how difficult it is to read the code. Again, this Halstead metric is formulated via the number operations and operands in code. This metric is used for a different perspective on code understanding.
	\item \textbf{Maintainability Index}, as defined in the Radon package\footnote{\url{https://radon.readthedocs.io/en/latest/intro.html\#maintainability-index}}, is a composite metric that is calculated using a number of other metrics as operands. We use it to assess how maintainable is the codebase.
\end{itemize}

\begin{figure*}[ht]
	\begin{subfigure}{.45\textwidth}
		\centering
		\includegraphics[width=\linewidth]{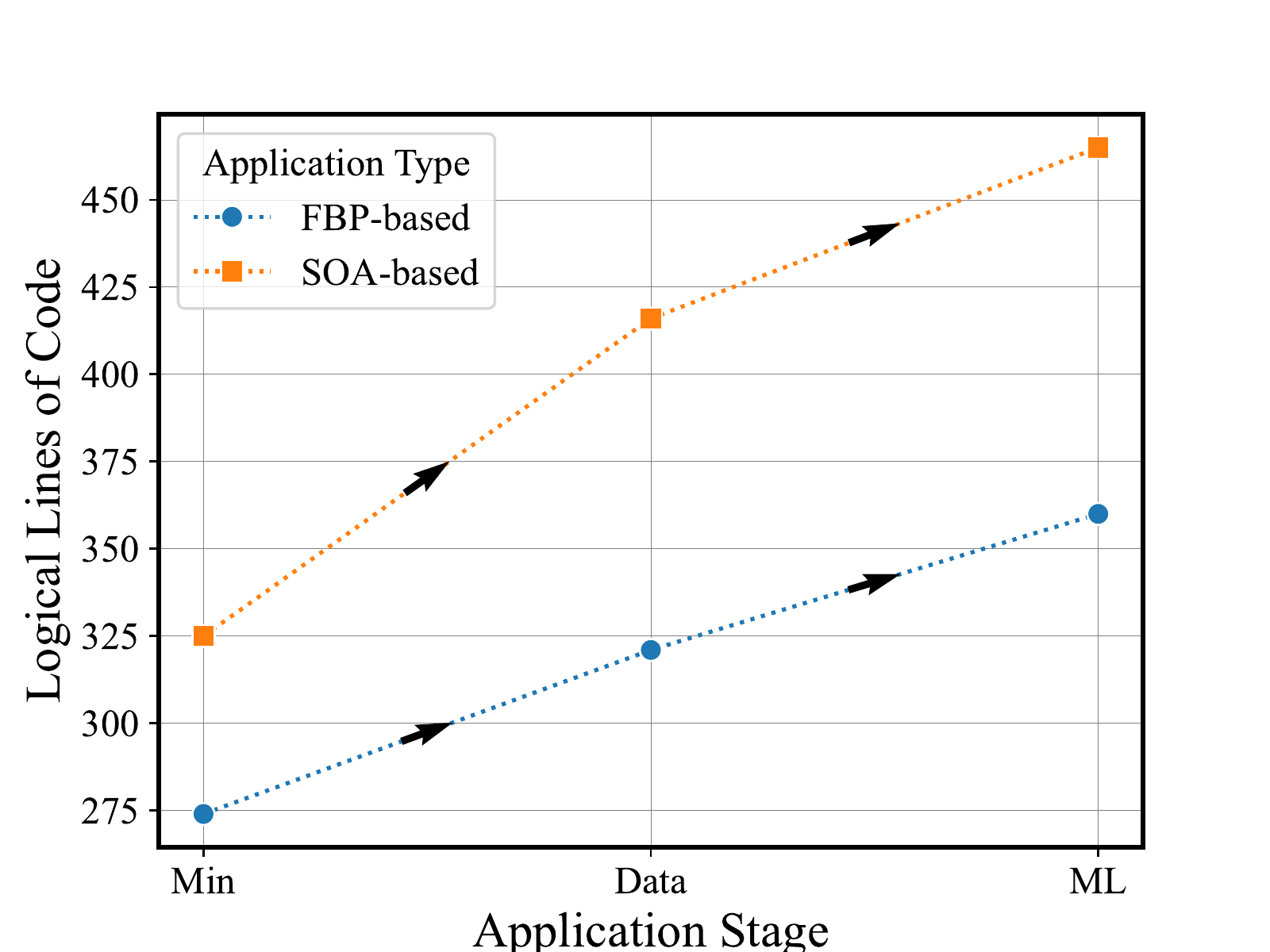}
	\end{subfigure}
	\begin{subfigure}{.45\textwidth}
		\centering
		\includegraphics[width=\linewidth]{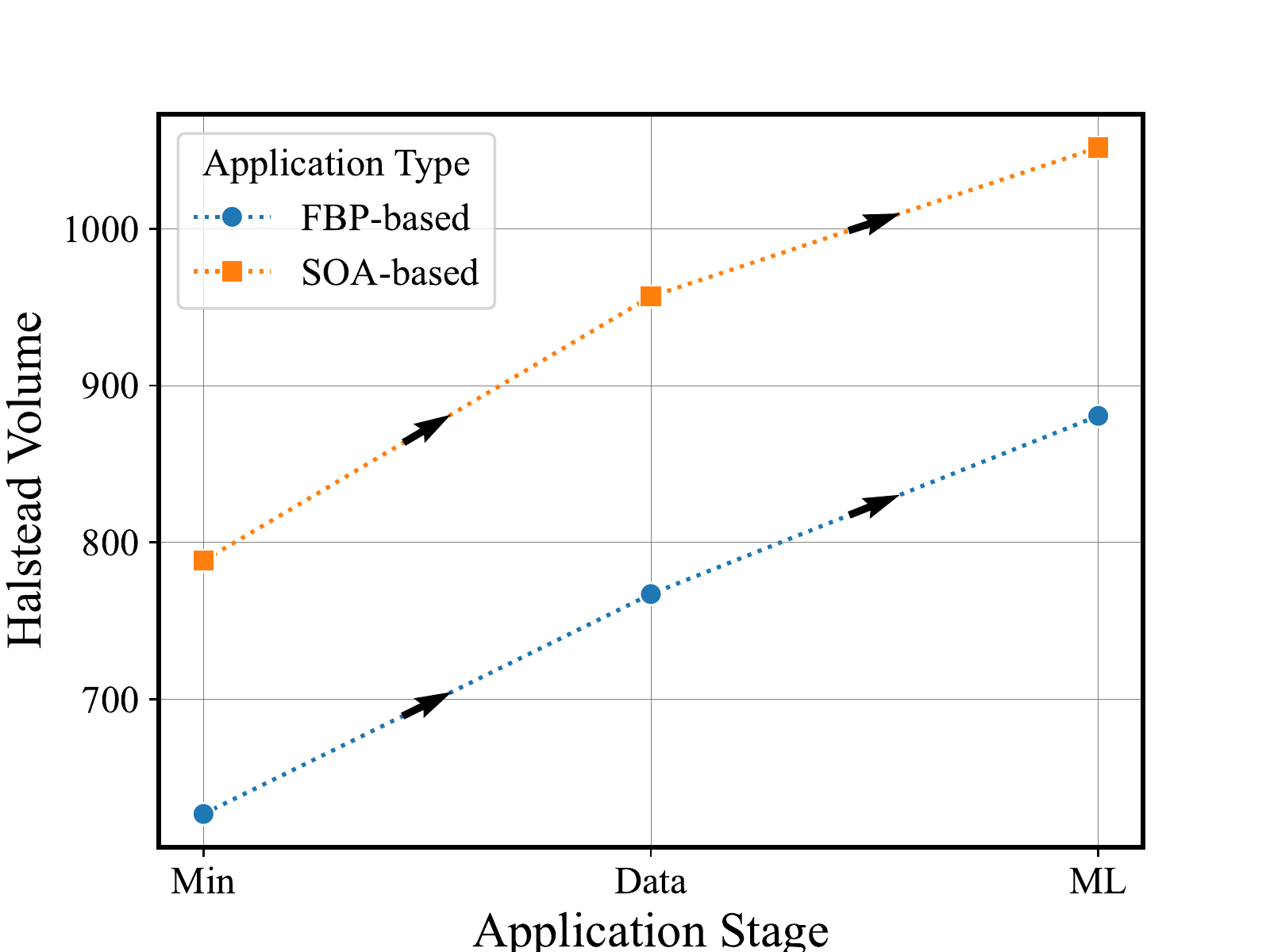}  
	\end{subfigure}
	\caption{Metrics to compare sizes of different implementations of the Ride Allocation app. Higher value corresponds to bigger codebase.}
	\label{figure:size_metrics}
\end{figure*}

\begin{figure}[ht]
	\centering
	\includegraphics[width=0.7\linewidth]{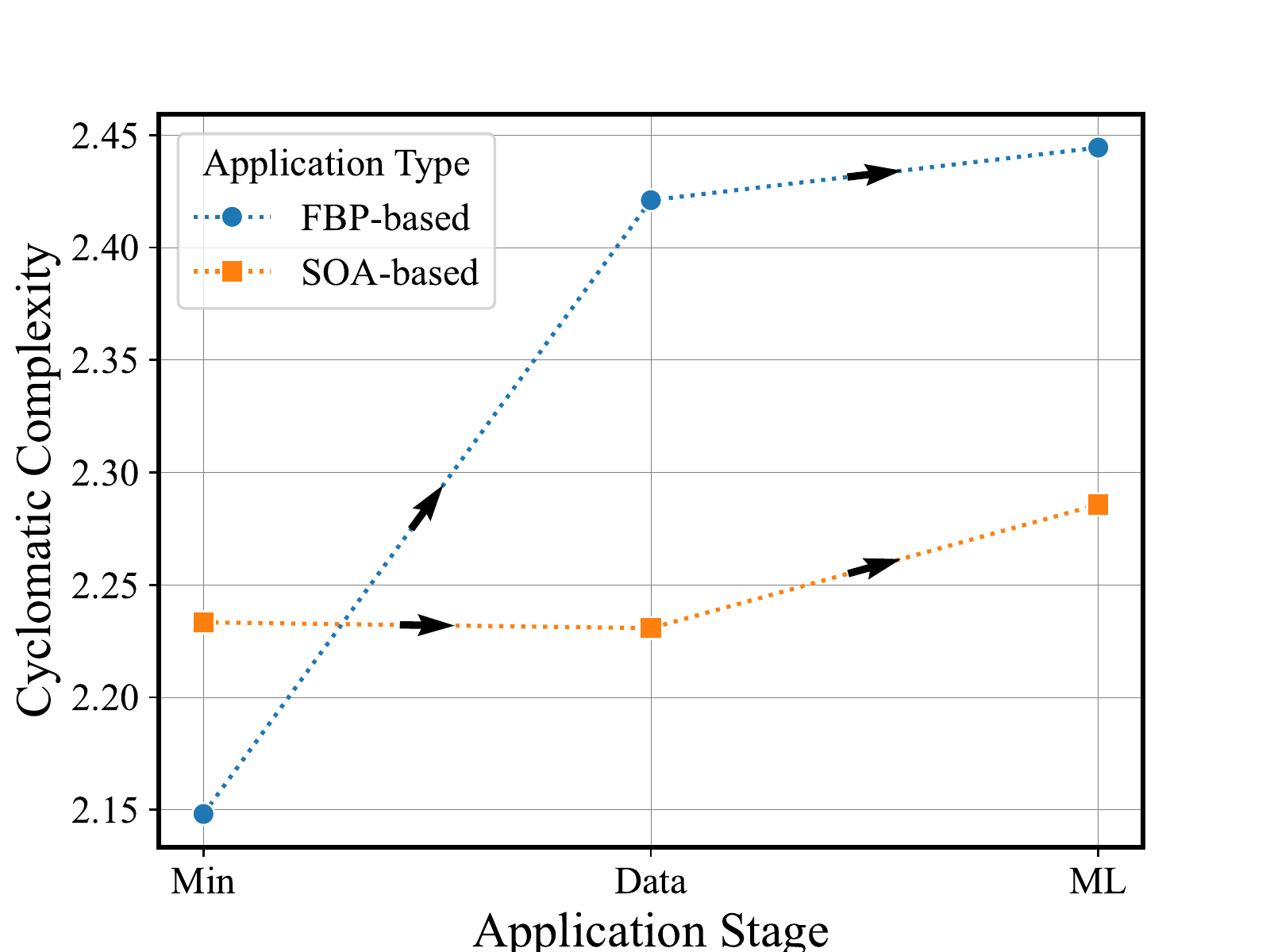}
	\caption{Comparison of average cyclomatic complexity of all implementations of the Ride Allocation app. Higher value means higher complexity.}
	\label{figure:cyclomatic_complexity}
\end{figure}

Figure~\ref{figure:size_metrics} presents the size comparison of all implementations. Despite being developed with different paradigms and different tools, the code of each implementation turned out to be of comparable size, both in logical lines of code and in terms of operations/operands. Across all stages in both paradigms the SOA implementation turned out to be marginally bigger. While this fact does not inform the comparison on its own, it simplifies further reasoning. If we had observed significant, say orders of magnitude, difference in size of the codebases, it would have complicated fair comparison of other metrics that scale with size, such as Halstead Difficulty.

The cyclomatic complexity metric comparison is presented on Figure~\ref{figure:cyclomatic_complexity}. It is clear that graph traversing logic that was required to collect the dataset in FBP-driven application has noticeable impact on the computational complexity of the solution. No other big increases are visible, indicating that no other changes either in FBP or in SOA codebase required complex logic to implement.

\begin{figure*}[ht]
	\begin{subfigure}{.45\textwidth}
		\centering
		\includegraphics[width=\linewidth]{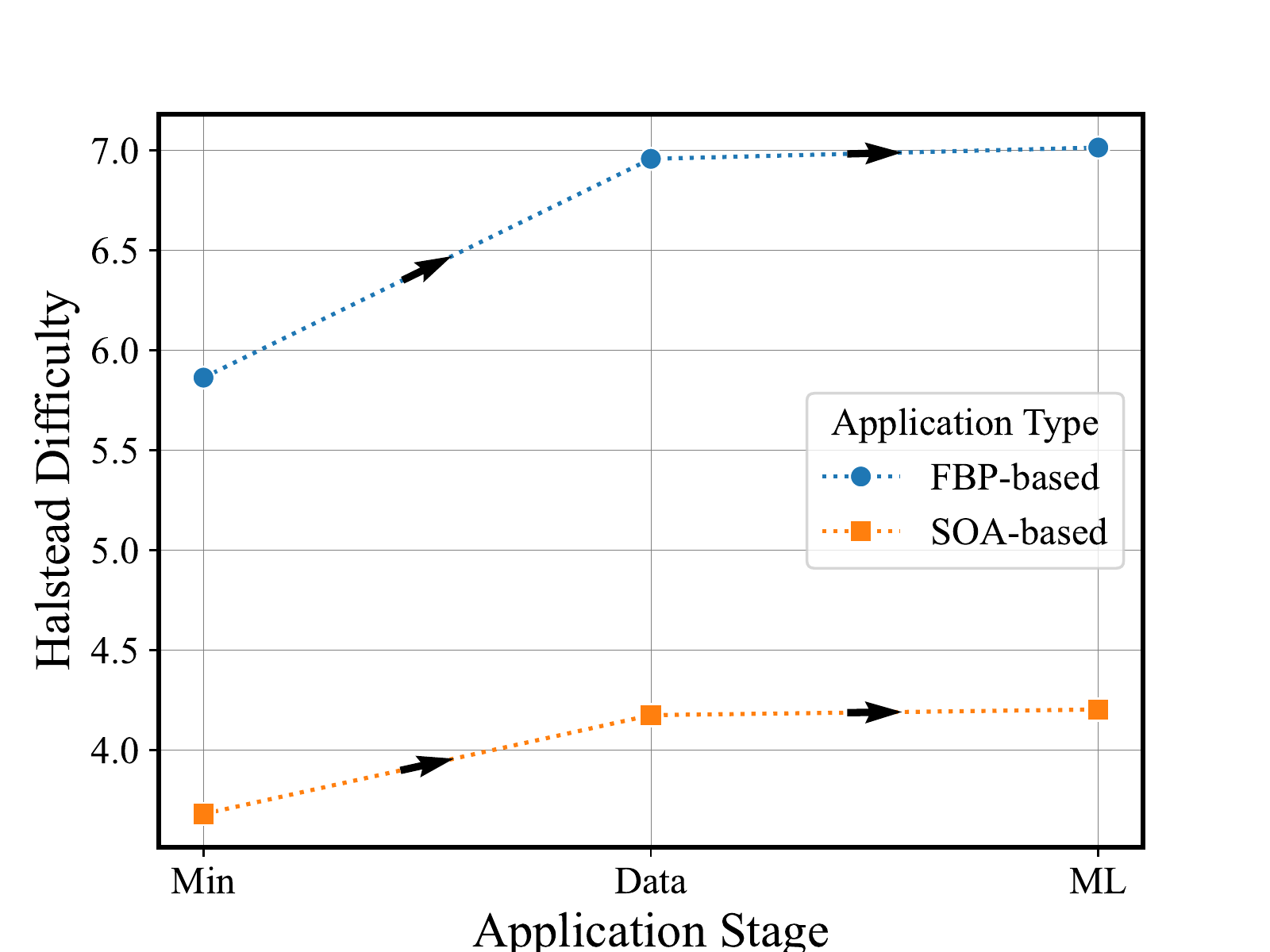}  
	\end{subfigure}
	\begin{subfigure}{.45\textwidth}
		\centering
		\includegraphics[width=\linewidth]{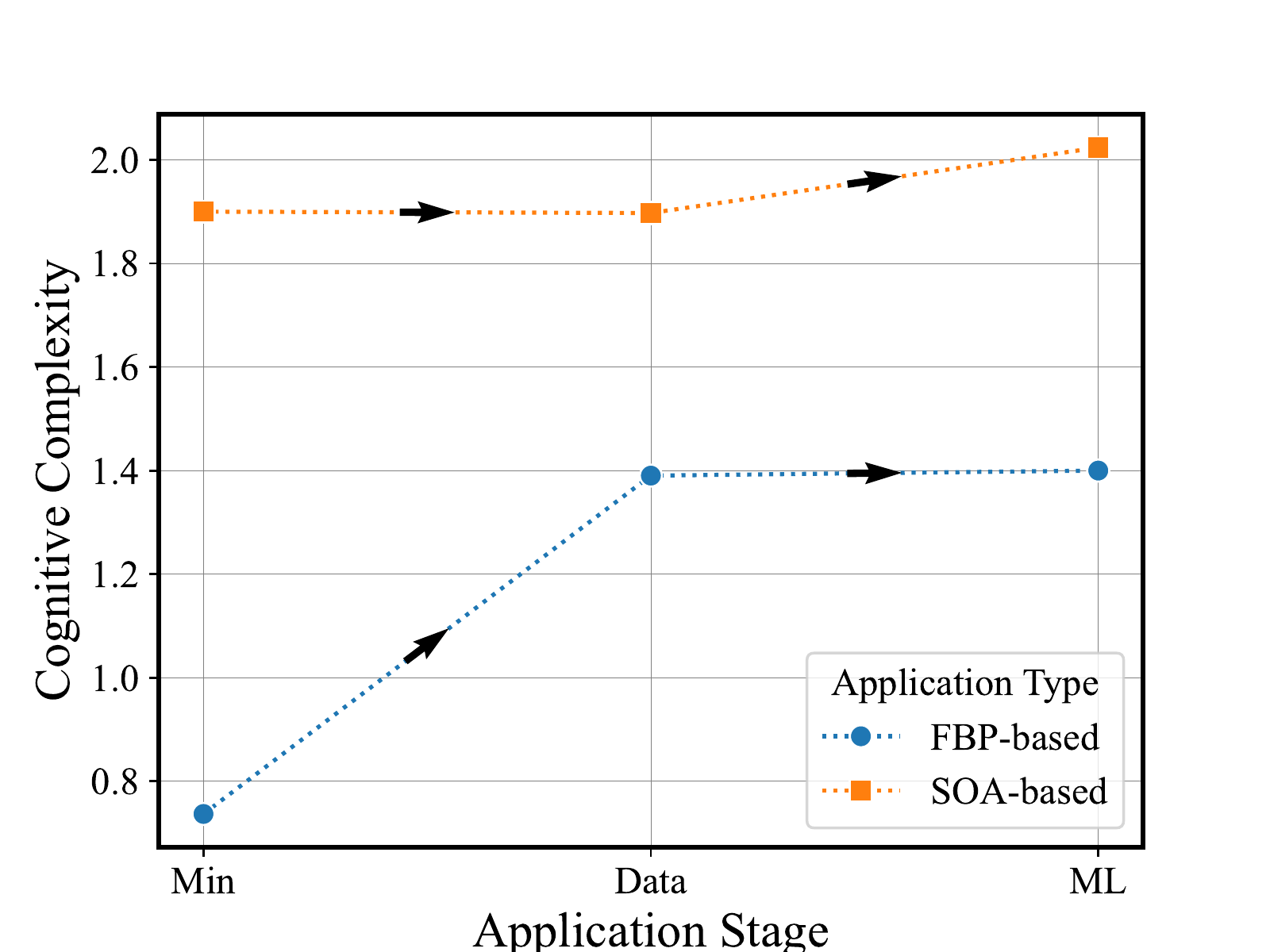}  
	\end{subfigure}
	\caption{Metrics to compare how difficult it is to understand and follow the code of different implementations of the Ride Allocation app. Average value of the cognitive complexity is used. For both metrics higher value means code is harder to comprehend.}
	\label{figure:cognitive_metrics}
\end{figure*}

Figure~\ref{figure:cognitive_metrics} presents two metrics that address understanding of the code by humans. It is interesting to observe that Halstead Difficulty and Cognitive Complexity metrics seemingly contradict: while the former shows SOA is a much simpler codebase, the latter clearly favors FBP. We believe this contradiction reflects key charasteristics of each development approach. FBP requires a lot of operations on edges and vertices in order to build and manipulate the data flow graph. That explains its worse score in Halstead Difficulty, which is calculated with operations and operands. However defining the graph is a linear process, and once the graph is defined control flow on it becomes trivial, thus FPB's lower score on Cognitive Complexity. On the contrary, SOA-based approaches require relatively fewer operations to setup, partially due to a richer software framework underlying them. However, the control flow of sending requests and handling responses (on top of the boilerplate already provided by Flask) isn't trivial, hence worse performance in terms of the Cognitive Complexity.

\begin{figure}[h]
	\centering
	\includegraphics[width=0.7\linewidth]{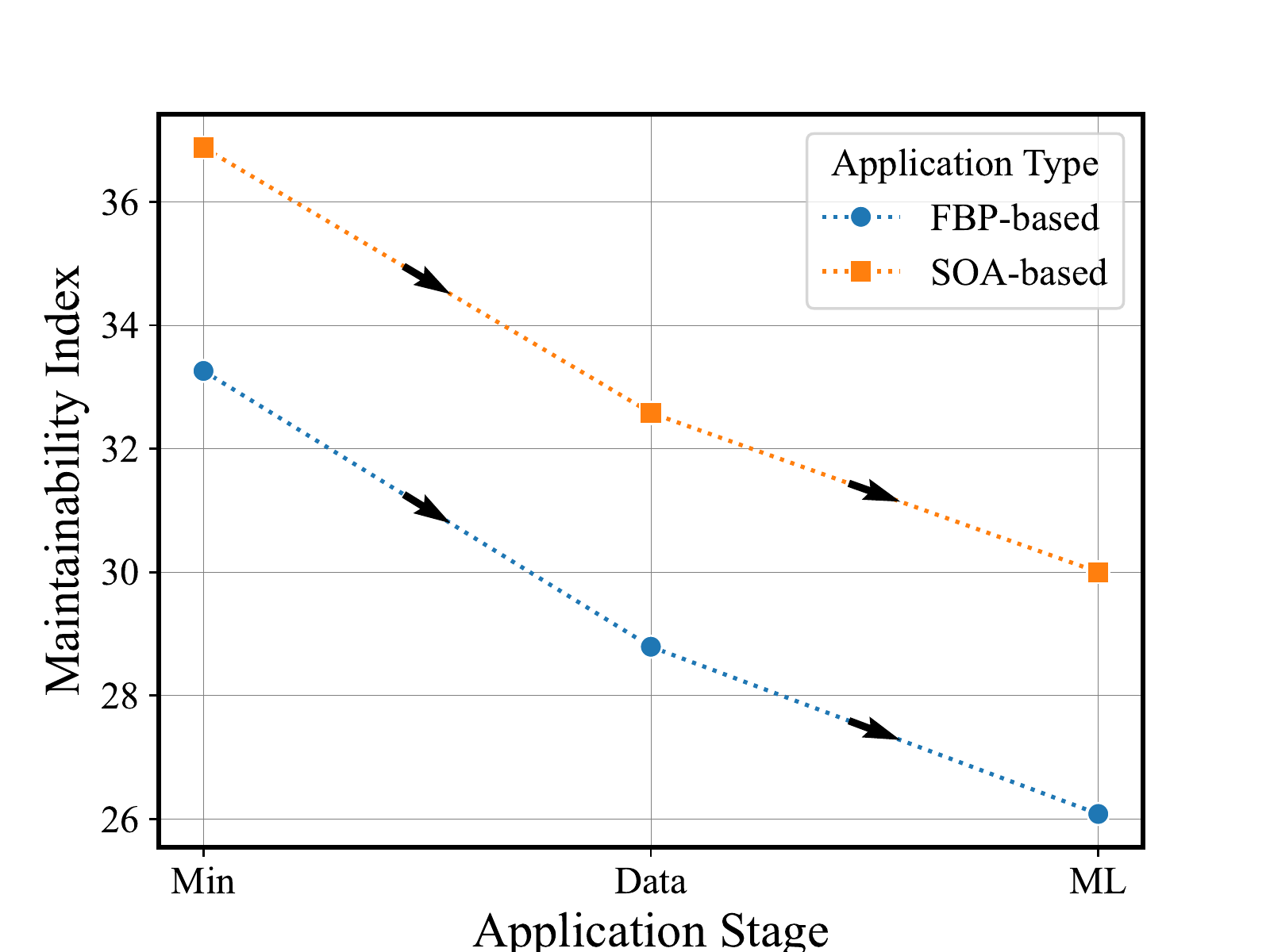}
	\caption{Comparison  of Maintainability Index metric of all implementations of the Ride Allocation app. Higher value means better maintainability.}
	\label{figure:maintainability_index}
\end{figure}

Nevertheless the Maintainability Index, as shown on the Figure~\ref{figure:maintainability_index}, is unaffected by aforementioned contradiction, and declines at approximately same rate for both paradigms.

Lastly, we combine multiple metrics together, to gain an additional insight. Figure~\ref{figure:combined_metrics} combines size-related metrics on one plot, and complexity-related metrics on the other. Interestingly, these plots make it very clear how data collection stage critically impacts all metrics of the codebase. For size metrics, the trends intersect between Stage 1 and Stage 2, while being nearly parallel between Stage 2 and Stage 3. This suggests that data collection on stage 2 had different scale of impact on different metrics for FBP and SOA implementations. For complexity metrics, the distance between points that represent Stage 1 and Stage 2 is significantly bigger than between points for Stage 2 and Stage 3. This again clearly shows how big was an impact on code complexity by the implementation of data collection.

\subsection{Empirical comparison}
Here we share more subjective impressions acquired in process of implementing the code for the experiment in both paradigms, that were hard to measure with objective metrics.

In Stage 1 the SOA application turned out quicker and more straightforward to develop than its FBP counterpart. SOA solution could be built with higher level tools, while FBP required some low level piping and additional abstractions that were not available out of the box. SOA paradigm is also more common, while FBP required a different way of thinking about the application logic. Overall, for someone who has experience with developing software services, the cost of developing Stage 1 code was lower for the SOA-based approach.

As expected, both approaches did not provide the dataset out of the box, and changes were necessary for dataset collection task in Stage 2. However the code needed for FBP solution was more localized. Essentially the whole data collection operation could be implemented at a single location. On the contrary, SOA implementation required changes in three distinct services in order to collect the complete dataset, which is more intrusive and prone to errors, especially if the code is ever anticipated to be in long term maintenance mode.

Both approaches performed well for deploying ML model in Stage 3: FBP exposed model predictions via separate processing node followed by an output stream, SOA encapsulated the model within a separate service with a simple API. We believe this has shown that both paradigms are well equipped for the usage of trained ML models.

\end{appendices}

\end{document}